\documentclass[onecolumn]{pasj00}

\SetRunningHead{B. Hatsukade \etal}{CO search from the host galaxy of GRB\,980425}

\title{
\large{
A Search for CO ($J$ = 3--2) Emission from the Host Galaxy of GRB\,980425 \\
with the Atacama Submillimeter Telescope Experiment}
}

\author{Bunyo \textsc{Hatsukade},\altaffilmark{1}
 		Kotaro \textsc{Kohno},\altaffilmark{1}
 		Akira \textsc{Endo},\altaffilmark{1,2}
 		Tomoka \textsc{Tosaki},\altaffilmark{3}
 		Kouji \textsc{Ohta},\altaffilmark{4}
		Seiichi \textsc{Sakamoto},\altaffilmark{2} \\
 		Nobuyuki \textsc{Kawai},\altaffilmark{5}
 		Juan R. \textsc{Cort\'{e}s},\altaffilmark{6,3}
 		Kouichiro \textsc{Nakanishi},\altaffilmark{3} 
 		Takeshi \textsc{Okuda},\altaffilmark{1,3}
 		Kazuyuki \textsc{Muraoka},\altaffilmark{1} \\
 		Takeshi \textsc{Sakai},\altaffilmark{2} 
 		Paul M. \textsc{Vreeswijk},\altaffilmark{7} 
 		Hajime \textsc{Ezawa},\altaffilmark{2}
 		Nobuyuki \textsc{Yamaguchi},\altaffilmark{2}
 		Kazuhisa \textsc{Kamegai},\altaffilmark{1} \\
 		and Ryohei \textsc{Kawabe}\altaffilmark{2}
 		}
 		
\altaffiltext{1}{Institute of Astronomy, the University of Tokyo, 2-21-1 Osawa, Mitaka, Tokyo 181-0015}
\email{hatsukade@ioa.s.u-tokyo.ac.jp}
\altaffiltext{2}{National Astronomical Observatory of Japan, 2-21-1 Osawa, Mitaka, Tokyo 181-8588}
\altaffiltext{3}{Nobeyama Radio Observatory, Minamimaki, Minamisaku, Nagano 384-1805}
\altaffiltext{4}{Department of Astronomy, Kyoto University, Kyoto 606-8502}
\altaffiltext{5}{Department of Physics, Tokyo Institute of Technology, 2-12-1 Ookayama, Meguro-ku, Tokyo 152-0033}
\altaffiltext{6}{Departamento de Astronom$\,\acute{\imath}$a, Universidad de Chile, Casilla 36-D, Santiago, Chile}
\altaffiltext{7}{European Southern Observatory, Casilla 19001, Santiago 19, Chile}
\altaffiltext{}{}
   
\Received{2006 August 1}
\Accepted{2006 November 21}
\SetVolumeData{2007}{59}{1}
\Published{2007 February 25}
\def\pageSTART{67}
\begin{document}
\maketitle

\begin{abstract}
We report on a deep search for $^{12}$CO ($J$ = 3--2) line emission from the host galaxy of GRB\,980425 with the Atacama Submillimeter Telescope Experiment (ASTE). 
We observed five points of the galaxy covering the entire region.
After combining all of the spectra, we obtained a global spectrum with the rms noise level of 3.3 mK in $T_{\mathrm{mb}}$ scale at a velocity resolution of 10 $\mathrm{km\ s^{-1}}$. 
No significant emission was detected, though we find a marginal emission feature in the velocity range corresponding to the redshift of the galaxy. 
We derive 3 $\sigma$ upper limits on the global properties: 
the velocity-integrated CO(3--2) intensity of $I_{\mathrm{CO}}$(3--2) $< 0.26\ \mathrm{K\ km\ s^{-1}}$ by adopting a velocity width of $67\ \mathrm{km\ s^{-1}}$; 
the H$_2$ column density of $N(\mathrm{H_2}) < 3 \times 10^{20} \mathrm{cm^{-2}}$; 
the molecular gas mass of $M(\mathrm{H_2}) < 3 \times 10^8\ \MO$, by assuming a CO line luminosity to H$_2$ molecular gas mass conversion factor of $X_{\mathrm{CO}} = 5.0\times10^{20}\ \mathrm{cm^{-2}\ (K\ km\ s^{-1})^{-1}}$; 
and the star formation rate of SFR $< 0.1\ \MO\ \mathrm{yr}^{-1}$, based on the Schmidt law.
The SFR is consistent with the previous results of H$\alpha$ and mid-IR observations, thereby suggesting that there is no significant obscured star formation in the host galaxy of GRB\,980425. 
This result implies that there is a variety of GRB hosts with regard to the presence of obscured star formation.
\end{abstract}

\section{Introduction}
Long-duration gamma-ray bursts (GRBs)---the most energetic event in the universe---are considered to be due to the death of massive stars (e.g., \cite{gala1998, stan2003}).
Therefore, GRBs are closely associated with the star formation of host galaxies. 
Since GRBs can be detected at cosmological distances (the current record is $z=6.3$ for GRB\,050904; \cite{kawa2006}), they are expected to be probes of the star formation history of the universe (e.g., \cite{tota1997, yone2004}).
In order to determine the use of GRBs, it is essential to understand the star-formation activity of their hosts. 

Multiwavelength observations have shown that GRB hosts are typically blue, sub-luminous dwarf galaxies (e.g., \cite{lefl2003}) with low metallicity (e.g., \cite{fynb2003}). The star formation rates (SFRs) determined from optical/UV observations are $\sim$1--10 $\MO\ \mathrm{yr}^{-1}$. On the other hand, submillimeter and radio continuum observations indicate that some GRB hosts are dusty and have massively star forming properties (SFR $\sim$ several 100 $\MO\ \mathrm{yr}^{-1}$) \citep{berg2003}. 
There is a wide discrepancy  between the SFRs derived from optical/UV and those from submillimeter/radio. 
Mid-IR observations further complicate the situation; only a small fraction of GRB hosts are detected at mid-IR wavelengths in contrast with the image of submillimeter/radio observations \citep{lefl2006}. 
These methods for deriving the SFRs each have their own drawbacks. 
Because optical/UV bands are subject to dust extinction and therefore the SFRs may be underestimated. 
Mid-IR, far-IR, and radio continuum are susceptible to contamination from active galactic nuclei (AGNs). 
The SFRs derived from submillimeter continuum have uncertainties in the assumption of the dust temperature and emission spectrum. 

In order to solve these problems, it is necessary to derive the SFRs in a method which is independent of existing methods and not affected by dust extinction and AGNs. 
For this purpose, an effective method is to observe the CO line. 
The CO line traces molecular gas, which is a fuel for star formation. 
Thus far, the search for CO ($J$ = 1--0) emission from the host galaxy of GRB\,030329 using the Nobeyama Millimeter Array \citep{kohn2005, endo2007} has been reported as the only attempt in this regard. 
However, only upper limits of the molecular gas mass and SFR of the host galaxy have been obtained. 
Since current instruments have limitations, it appears that target selection is essential to detect CO line emission. 

\begin{figure}
	\begin{center}
    	\FigureFile(80mm,71mm){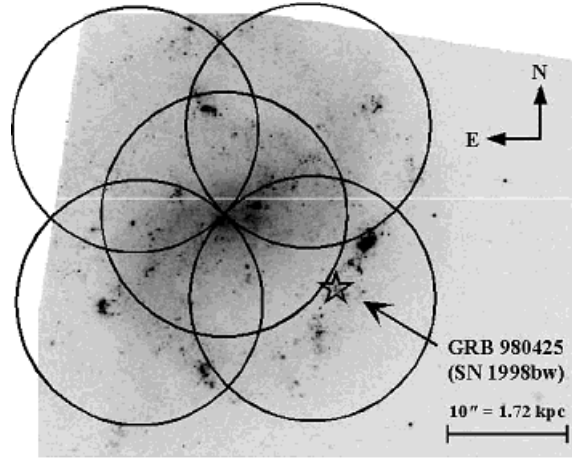}
	\end{center}
	\caption{Hubble Space Telescope image \citep{fynb2000} and ASTE observation beams (circles). The HPBW of 22$''$ corresponds to 3.8 kpc at the distance of the galaxy. The position of the GRB is marked with a star.}
	\label{fig:grbhost}
\end{figure}

In this paper, we report on observations of the $J$ = 3--2 transition line of $^{12}$CO in the host galaxy of GRB\,980425 using the Atacama Submillimeter Telescope Experiment (ASTE: \cite{kohn2004, ezaw2004}). 
GRB\,980425 emerged in an H\emissiontype{II} region in a spiral arm and was identified with SN\,1998bw (type Ic supernova). 
Its isotropic $\gamma$-ray energy of $\sim$$8 \times 10^{47}$ erg is significantly less than that of typical GRBs by several orders of magnitude \citep{gala1998}.
The host galaxy termed ESO184-G82 (figure \ref{fig:grbhost}) is the nearest GRB host known to date. The redshift of $z=0.0085 \pm 0.0002$ \citep{tinn1998} is extremely low among GRB hosts (the mean redshift is $z = 2.8$; \cite{jako2006}). 
Due to the proximity, it is the best target to detect CO line emission.
The host galaxy is a blue, sub-luminous, and low-metallicity star-forming galaxy (see table \ref{tab:grbhost}). 
SFR is derived from previous H$\alpha$, mid-IR, and hard X-ray observations as $0.35\ \MO\ \mathrm{yr}^{-1}$ \citep{soll2005}, $0.4\ \MO\ \mathrm{yr}^{-1}$ \citep{lefl2006}, and $2.8 \pm 0.3\ \MO\ \mathrm{yr}^{-1}$ \citep{wats2004} respectively. 
The discrepancy between the SFRs derived from the X-ray and H$\alpha$ observations implies that the host galaxy contains a large amount of molecular gas.
The detection of CO line emission is simplified by the fact that the galaxy has one of the highest values of metallicity---[log(O/H)+12] = 8.6 \citep{soll2005}---among GRB hosts (e.g., \cite{stan2006}).

We assume a cosmology with $H_0=71\ \mathrm{km\ s^{-1}\ Mpc^{-1}}$, $\Omega_{\mathrm{M}}=0.27$, and $\Omega_{\Lambda}=0.73$. 
The luminosity distance of GRB\,980425 is $D_{\mathrm{L}}=36.1$ Mpc and the angular distance is $D_{\mathrm{A}}=35.5$ Mpc ($1''$ corresponds to 0.172 kpc).

\begin{table}
\begin{center}
\caption{Properties of the host galaxy of GRB\,980425}
\label{tab:grbhost}
\begin{tabular}{cccccccc}
 \hline\hline
 R.A. & Decl.& $z$ & $i$ & $M_B$ & $L(\mathrm{H\alpha})$ & $L(\mathrm{IR})$ 
 & $\log{(\mathrm{O/H})}+12$ \\
 (J2000.0) & (J2000.0) &  & (degree) & (mag) & $(\mathrm{erg\ s^{-1}})$ & $(\LO)$ \\
 (1) & (2)  & (3) & (4)   & (5)                      & (6)     & (7) & (8) \\
 \hline
 \timeform{19h35m04s.4} \altaffilmark{a} 
 & \timeform{-52D50'38''} \altaffilmark{a} 
 & $0.0085 \pm 0.0002$ \altaffilmark{b} 
 & 30 \altaffilmark{a}
 & --17.65 \altaffilmark{c} 
 & $4.4\times 10^{40}$ \altaffilmark{c} 
 & $2\times 10^9$ \altaffilmark{d} 
 & 8.6 \altaffilmark{c} \\
 \hline 
 \multicolumn{8}{@{}l@{}}{\hbox to 0pt{\parbox{180mm}{\footnotesize
 \vspace{2mm}
   (1) Right ascension. (2) Declination. 
   (3) Redshift.
   (4) Inclination.
   (5) Absolute magnitude. 
   (6) H$\alpha$ luminosity. \\
   (7) IR luminosity determined by using the correlations between mid-IR luminosity and 8--1000 $\micron$ integrated emission. 
   (8) Metallicity.
   \par\noindent
   \altaffilmark{a} 2MASS; 
   \altaffilmark{b} \cite{tinn1998}; 
   \altaffilmark{c} \cite{soll2005};
   \altaffilmark{d} \cite{lefl2006}.
 }\hss}}
\end{tabular}
\end{center}
\end{table}

\section{Observations and data reduction}

Observations of $^{12}$CO ($J$ = 3--2) were conducted using the ASTE on June 14--22, 2005.
The ASTE is a single-dish 10 m submillimeter telescope at Pampa la Bola, Chile, equipped with a 4 K cooled superconductor-insulator-superconductor (SIS) mixer receiver (\cite{seki2001, mura2007}).
The observations were carried out remotely from ASTE operation rooms in San Pedro de Atacama, Chile, and in Mitaka, Japan, by using the network observation system N-COSMOS3, developed by the National Astronomical Observatory of Japan (NAOJ) \citep{kama2005}. 
The observation frequency was set to the redshifted $^{12}$CO ($J$ = 3--2) line of 342.882 GHz at the upper side band (USB). The half-power beam width (HPBW) was 22$''$, corresponding to 3.8 kpc at the distance of the galaxy.
We used 4 digital spectrometers (A1, A2, A3, and A4) with a bandwidth of 512 MHz and 1024 channels \citep{sora2000}. 
A1 and A2 were configured at the center of the USB, and A3 and A4 were moved from the center by $\mp 256$ MHz respectively to cover a bandwidth of 1024 MHz ($\sim$$900\ \mathrm{km\ s^{-1}}$).
The weather was good during the observations and the system temperature was in the range of 250--750 K in DSB.
In order to cover the entire region of the galaxy, we observed 5 points around the host galaxy center with a spacing of 11$''$, including the GRB position (see figure \ref{fig:grbhost}).
The sky emission was subtracted by position switching. 
The pointing was checked every few hours using the CO(3--2) emission of W Aql and continuum emission of Mars, and the accuracy was within 3$''$. 
The intensity calibration was performed by the chopper wheel method.
The absolute intensity calibration was performed by observing the CO(3--2) emission of M17SW and assuming that the velocity-integrated CO(3--2) intensity, $I_{\mathrm{CO}}$(3--2), is $536.3\ \mathrm{K\ km\ s^{-1}}$ \citep{wang1994}. 
The main beam efficiency was measured to be 0.59--0.64 during the observations, and we adopted a constant value of 0.62.

Data reduction was carried out using NEWSTAR GBASE developed by the Nobeyama Radio Observatory (NRO).
We used only data when the system temperature was less than 500 K. 
The total integration time was about 3 hours at the host galaxy center and about 1.5 hours at the other points.
Linear baselines were subtracted from each spectrum.
The rms noise level decreased with the square root of time, thereby indicating that the system was sufficiently stable for long integration.

\section{Results}
\begin{figure}
	\begin{center}
    	\FigureFile(114mm,107mm){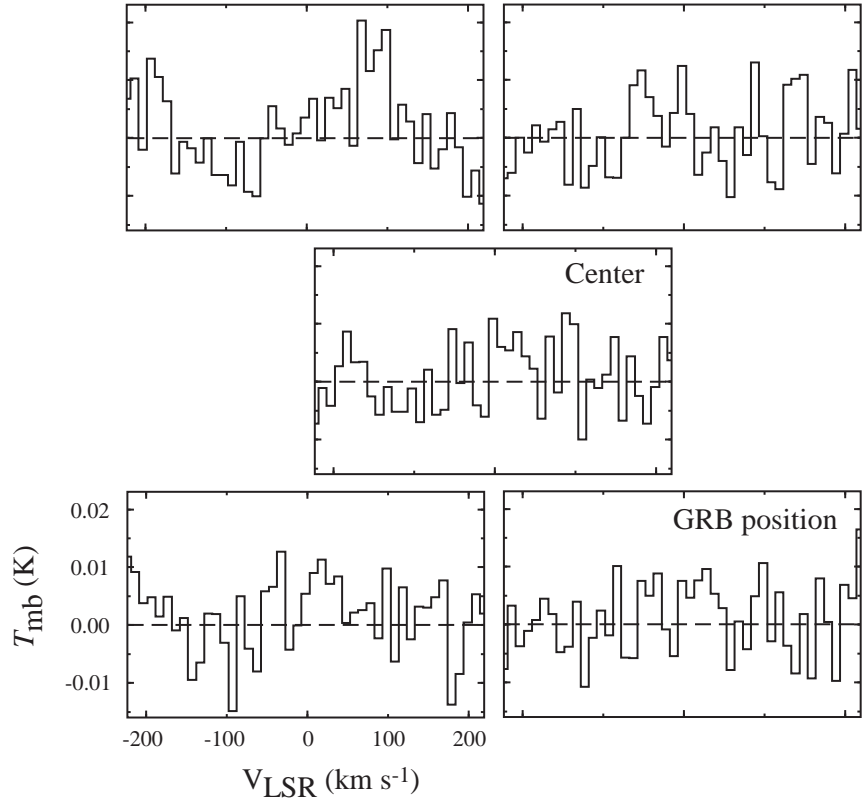}
	\end{center}
	\caption{Spectra of five observation points at a velocity resolution of 10 km s$^{-1}$. The peak main beam temperatures are $\sim$13--20 mK. The rms noise levels are $\sim$6 mK in $T_{\mathrm{mb}}$ scale.}
	\label{fig:grb5p}
\end{figure}

\begin{figure}
	\begin{center}
    	\FigureFile(142mm,49mm){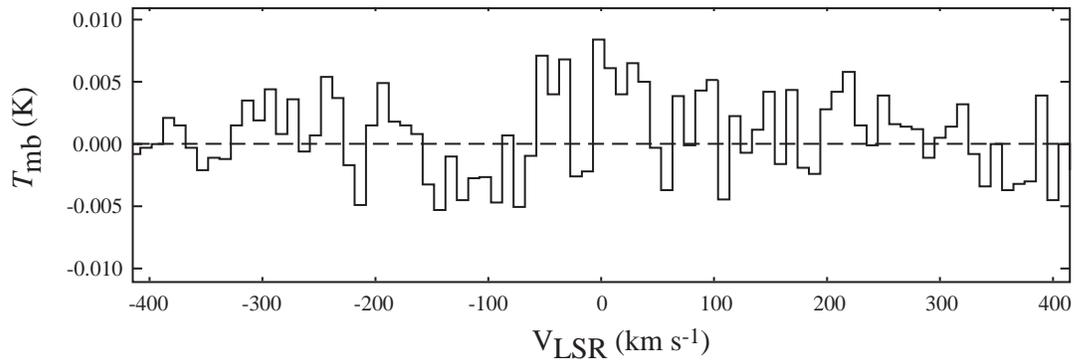}
	\end{center}
	\caption{Global spectrum at a velocity resolution of 10 km s$^{-1}$. 
	This exhibits the global property of the host galaxy of GRB\,980425. 
	The rms noise level is 3.3 mK in $T_{\mathrm{mb}}$ scale.}
	\label{fig:grb1p}
\end{figure}

The spectra of 5 points using the data of spectrometer A1 and A2 are shown in figure \ref{fig:grb5p}. 
The rms noise levels are $\sim$6 mK in the scale of main beam temperature ($T_{\mathrm{mb}}$) at a velocity resolution of 10 $\mathrm{km\ s^{-1}}$. 
By combining all spectra of 4 spectrometers, we obtained a global spectrum covering $\sim$800 $\mathrm{km\ s^{-1}}$ with the rms noise level of 3.3 mK in $T_{\mathrm{mb}}$ scale at a velocity resolution of 10 $\mathrm{km\ s^{-1}}$ (figure \ref{fig:grb1p}). 
When combining, we cut out 10\% of the bandwidth of both ends, since the sensitivity is low. 
No significant emission is detected, although there seems to be an emission feature at the center of the global spectrum. 
More observations are needed to confirm this possibility. 

We now discuss the 3 $\sigma$ upper limits on the global properties of the host galaxy from now on. 
In order to estimate an upper limit of $I_{\mathrm{CO}}$(3--2), we assume the velocity width of the galaxy. 
In the sample of \citet{lero2005}, rotation velocities of 115 dwarf galaxies ($M_B \gtrsim -18$) are in the range 21--150 $\mathrm{km\ s^{-1}}$ and the average is $67\ \mathrm{km\ s^{-1}}$. 
Correcting for the inclination of the host galaxy multiplying the velocity by 2, we assume the velocity width of $67\ \mathrm{km\ s^{-1}}$. 
The rms of the global spectrum at the this velocity resolution is 1.3 mK, and therefore the 3 $\sigma$ upper limit of $I_{\mathrm{CO}}$(3--2) is $0.26\ \mathrm{K\ km\ s^{-1}}$.

\section{Discussion}

\subsection{H$_2$ Column Density}
The 3 $\sigma$ upper limit of H$_2$ column density is calculated to be $3 \times 10^{20}\ \mathrm{cm^{-2}}$ as follows:
\begin{eqnarray}
N(\mathrm{H_2})
= X_{\mathrm{CO}} \cdot I_{\mathrm{CO}}(3-2) \cdot R_{32/10}^{-1} \cdot \cos{(i)}\ , 
\end{eqnarray}
where $X_{\mathrm{CO}}$ is a CO luminosity to H$_2$ molecular gas mass conversion factor in units of $\mathrm{cm^{-2}\ (K\ km\ s^{-1})^{-1}}$, 
$R_{32/10}$ is a CO(3--2)/CO(1--0) integrated line intensity ratio,
and $i$ is the inclination of the disk. 
A conversion factor, $X_{\mathrm{CO}}$, is derived using the correlation between $X_{\mathrm{CO}}$ and the metallicity \citep{arim1996}. 
By using the metallicity of the host galaxy of $[12+\log(\mathrm{O/H)}]=8.6$, $X_{\mathrm{CO}}$ is estimated as $5.0\times10^{20}\ \mathrm{cm^{-2}\ (K\ km\ s^{-1})^{-1}}$. 
$R_{32/10}$ is in the range of 0.2--1.2 depending on the type of the galaxy such as nearby dwarf starburst galaxies, early-type galaxies, starburst spiral galaxies, luminous infrared galaxies (LIRGs), and ultraluminous infrared galaxies (ULIRGs) \citep{meie2001, deve1994, maue1999, vila2003, yao2003, nara2005}. 
In this paper we adopt $R_{32/10}=0.4$, the typical value of the Galactic disk \citep{sand1993}.

\subsection{Molecular Gas Mass}
The 3 $\sigma$ upper limit of molecular gas mass of $M(\mathrm{H_2}) < 3 \times10^8\ \MO$ is obtained from
\begin{eqnarray}
M(\mathrm{H_2})
= \Sigma_{\mathrm{H_2}} \cdot S \cdot \cos{(i)}^{-1}\ ,  
\end{eqnarray}
where $\Sigma_{\mathrm{H_2}}$ is the H$_2$ surface density, and $S$ is the area of the total beam, which is subtended by the 5 observation beams. 
This is consistent with those of dwarf galaxies \citep{lero2005, meie2001}.

\subsection{Star Formation Rate}
We derive the 3 $\sigma$ upper limit of star formation rate of $0.1\ \MO\ \mathrm{yr}^{-1}$ by applying the Schmidt law \citep{kenn1998}:
\begin{eqnarray}
\mathrm{SFR}
= 2.5\times 10^{-4} \cdot (\Sigma_{\mathrm{H_2}})^{1.4} \cdot S \cdot \cos{(i)}^{-1} \ \MO\ \mathrm{yr}^{-1} \ . 
\end{eqnarray}
This is consistent with the results of the previous H$\alpha$ observations (SFR = $0.35\ \MO\ \mathrm{yr}^{-1}$, \cite{soll2005}) considering the uncertainties stated above.
This indicates that the host galaxy has no significant obscured star formation.
This is also consistent with the value of mid-IR observations---SFR = $0.4\ \MO\ \mathrm{yr}^{-1}$ \citep{lefl2006}.

\begin{figure}
	\begin{center}
    	\FigureFile(113mm,129mm){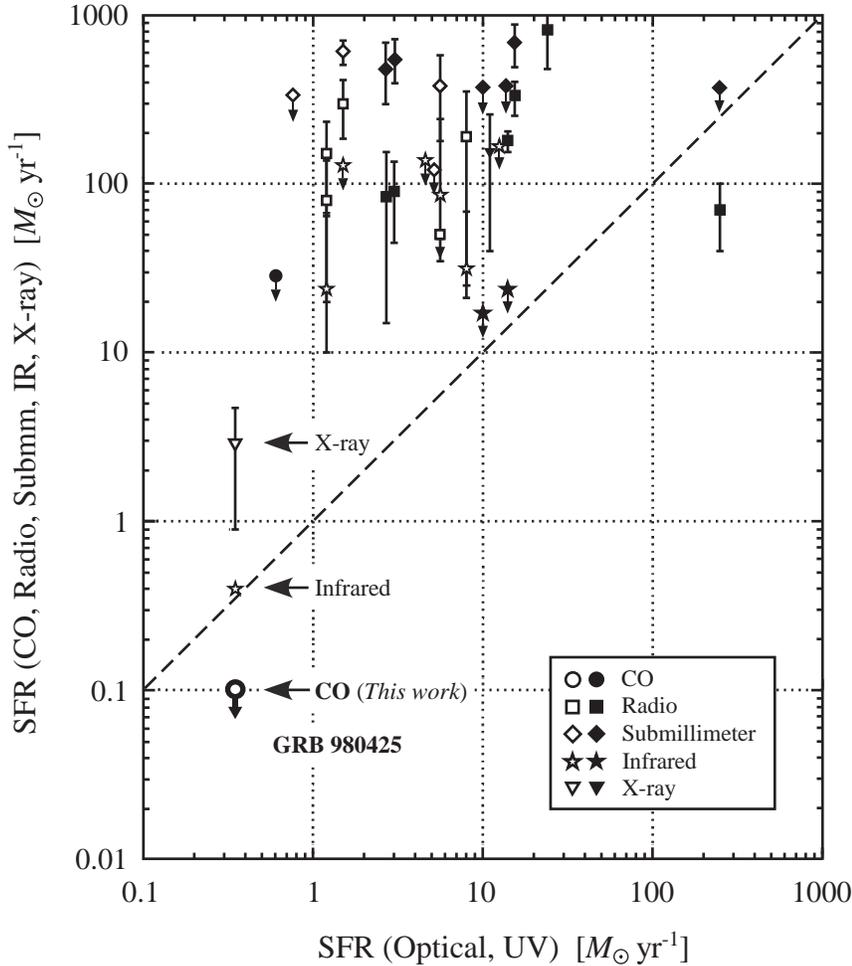}
	\end{center}
	\caption{Comparison of the SFRs of GRB hosts determined by various observational methods. The ordinate is the SFR derived by extinction-free methods such as the CO line, radio continuum, submillimeter continuum, infrared continuum, and X-rays. The abscissa is the SFR from optical (recombination and forbidden lines) and UV continuum. The data are based on \citet{endo2007}. The open symbols are the SFRs that are corrected for extinction in the host galaxies, and the solid symbols are those that are not corrected. The down-pointing arrows represent upper limits. The diagonal indicates equal values for both axes. It is clear that most of the GRB hosts are above this line, that is, the SFRs from the extinction-free methods are higher than those from the optical/UV bands even after extinction correction. On the other hand, the host galaxy of GRB\,980425 show a different trend.}
	\label{fig:sfr}
\end{figure}

Figure \ref{fig:sfr} shows the SFRs of GRB hosts derived by several methods (see \cite{endo2007} and references therein). 
The ordinate is the SFR determined from extinction-free wavelengths, such as the CO line, radio continuum, submillimeter continuum, infrared continuum, and X-rays.
The abscissa is the SFR determined from optical lines (recombination and forbidden lines) and UV continuum. 
The diagonal indicates that the values of both axes are equal. 
The majority of the GRB hosts are located above this line, implying that they have a large amount of molecular gas and massive star formation obscured by dust. 
This tendency is observed in LIRGs, ULIRGs, and submillimeter galaxies but not in normal spiral galaxies (e.g., \cite{youn1996, berg2003}). 
On the other hand, our study shows that the host galaxy of GRB\,980425 shows a different trend. 
This suggests that various GRB hosts exist in terms of the presence of obscured star formation. 

\begin{table*}
\begin{center}
\caption{3 $\sigma$ upper limits of the global properties of the host galaxy}
\label{tab:physical}
\begin{tabular}{ccccc}
 \hline\hline
   $I_{\mathrm{CO}}$(3--2) 
 & $N(\mathrm{H_2})$ 
 & $\Sigma_{\mathrm{H_2}}$ 
 & $M(\mathrm{H_2})$  
 & SFR    \\
 ($\mathrm{K\ km\ s^{-1}}$)  
 & ($\mathrm{cm}^{-2}$) 
 & ($\MO$ pc$^{-2}$)
 & ($\MO$) 
 & ($\MO$ yr$^{-1}$) \\
 (1)             & (2)       & (3)        & (4)          &(5)  \\
 \hline 
 0.26 & $3\times 10^{20}$ &  5  &  $3 \times 10^8$  & 0.1 \\  
 \hline
 \multicolumn{5}{@{}l@{}}{\hbox to 0pt{\parbox{125mm}{\footnotesize
 \vspace{1mm}
   (1) Velocity-integrated CO(3--2) line intensity. 
   (2) H$_2$ column density. \\
   (3) H$_2$ surface density. 
   (4) Molecular gas mass. 
   (5) Star formation rate.
   \par\noindent
 }\hss}}
\end{tabular}
\end{center}
\end{table*}

\section{Summary}
We searched for $^{12}$CO ($J$ = 3--2) line emission from the host galaxy of GRB\,980425 with the ASTE. 
Five points were observed around the host galaxy center covering the whole area.
The results are as follows:
\begin{enumerate}
	\item No significant emission is detected, but there seems to be a marginal emission feature in the velocity range corresponding to the redshift of the galaxy. 
	After combining all spectra, we obtained a global spectrum with the rms noise level of 3.3 mK in $T_{\mathrm{mb}}$ scale at a velocity resolution of 10 $\mathrm{km\ s^{-1}}$.

	\item The 3 $\sigma$ upper limit of velocity-integrated CO(3--2) intensity is $I_{\mathrm{CO}}$(3--2) $< 0.26\ \mathrm{K\ km\ s^{-1}}$ by assuming a velocity width of $67\ \mathrm{km\ s^{-1}}$. 
	
	\item The 3 $\sigma$ upper limits of H$_2$ column density and molecular gas mass are $N(\mathrm{H_2}) < 3 \times 10^{20}\ \mathrm{cm^{-2}}$ and $M(\mathrm{H_2}) < 3 \times 10^8\ \MO$ respectively 
	by adopting a CO(3--2)/CO(1--0) integrated line intensity ratio of $R_{32/10}=0.4$ and a CO luminosity to H$_2$ molecular gas mass conversion factor of $X_{\mathrm{CO}} = 5.0 \times 10^{20}\ \mathrm{cm^{-2}\ (K\ km\ s^{-1})^{-1}}$. 
	These are consistent with the values of nearby dwarf galaxies. 
 
	\item The 3 $\sigma$ upper limit of star formation rate is SFR $< 0.1 \ \MO\ \mathrm{yr}^{-1}$, based on the Schmidt law. 
	This is consistent with the results of H$\alpha$ and mid-IR observations, suggesting that there is no significant obscured star formation in the host galaxy of GRB\,980425. This result implies that various GRB hosts exist in terms of the presence of obscured star formation. 
\end{enumerate}

We would like to acknowledge the members of the ASTE team for the operation and ceaseless efforts to improve the ASTE. This study was financially supported by the MEXT Grant-in-Aid for Scientific Research on Priority Areas No. 15071202.

\end{document}